\documentclass{PoS}
\usepackage{amssymb}
\usepackage{latexsym}
\usepackage{amsmath}
\usepackage{slashed}
\usepackage{graphicx} 
\usepackage{subfigure}

\title{Flavor violating signatures of neutral Higgs bosons at the LHeC}

\ShortTitle{Flavor violating signatures of neutral Higgs bosons}

\author{\speaker{J. Hern\'andez-S\'anchez}\thanks{Supported by SNI-CONACYT and PROMEP-SEP grants.}\\
        Facultad de Ciencias de la Electr\'onica, Benem\'erita Universidad Aunt\'onoma de Puebla, Apdo. Postal 542, C.P. 72570 Puebla, Puebla, M\'exico.\\
 Dual C-P Institute of High Energy Physics, Puebla, Pue., M\'exico.\\
        E-mail: \email{jaime.hernandez@correo.buap.mx}}

\author{S. P. Das\\
        Instituto de F\'{\i}sica, Benem\'erita Universidad Aut\'onoma de Puebla, Apdo. Postal J-28, C.P. 72570 Puebla, Puebla, M\'exico.\\
        E-mail: \email{sprasad@ifuap.buap.mx}}

\author{S. Moretti \\
School of Physics and Astronomy, University of Southampton, Highfield, Southampton SO17 1BJ, United Kingdom and Particle Physics Department, Rutherford Appleton Laboratory, Chilton, Didcot, Oxon OX11 0QX, United Kingdom \\
E-mail: \email{s.moretti@soton.ac.uk}} 
       
\author{Alfonso Rosado\\
        Instituto de F\'{\i}sica, Benem\'erita Universidad Aut\'onoma de Puebla, Apdo. Postal J-28, C.P. 72570 Puebla, Puebla, M\'exico.\\
        E-mail: \email{rosado@ifuap.buap.mx}}
        
\author{Reyna Xoxocotzi-Aguilar\\
      Facultad de Ciencias F\'{\i}sico-Matem\'aticas,Benem\'erita Universidad Aut\'onoma de Puebla, Apdo. Postal 1364, C.P. 72570 Puebla, Puebla, M\'exico.\\
        E-mail: \email{xoxo\_reyna@yahoo.com.mx}}

\abstract{We analyze the prospect for observing the lightest and heaviest CP-even neutral Higgs bosons ($\phi=h,\;H$) in their decays to flavor violating $b\bar{s}$ (with charge conjugate) final states at the future Large Hadron electron Collider (LHeC), in the framework of the Two Higgs Doublet Model type III (2HDM-III) assuming a four-zero texture in the Yukawa matrices and a general Higgs potential. We consider the charge current production processes $e^- p \to \nu_e \phi q_f$ ($\phi= h, \; H$), with the flavor violating decays of the Higgs bosons. We consider scenarios of the model which are consistent with  current experimental data from flavor and Higgs physics. We also consider the most relevant Standard Model (SM) backgrounds. We find that the lightest (SM-like) CP-even Higgs boson $h$ and the heaviest one $ H$ (with mass up to 150 GeV) would have 1--2 $\Sigma$ and 1 $\Sigma$ significances, respectively, with 100 fb$^{-1}$ of data.  At the end of the LHeC running, the significances would be increased by ten times and the detection of both Higgs bosons ($h, H$) would be guaranteed.}

\FullConference{XXIII International Workshop on Deep-Inelastic Scattering,\\
		27 April - May 1 2015\\
		Dallas, Texas}

\begin{document}
\section{Introduction}
 Since the Higgs boson has been discovered at the Large Hadron Collider (LHC) by  the ATLAS \cite{Atlas} and CMS 	\cite{cms} experiments, the Standard Model (SM) is well established. However, recently CMS reported a slight excess of  Lepton Flavor-Violating (LFV) Higgs decays $h \to \mu^\mp \tau^\pm$ \cite{Khachatryan:2015kon}, which is forbidden in the SM and may be a hint of new physics. However, this is not an indication of which new physics may be at work, as one can explain such flavor-violating processes  in severals extensions of the Higgs sector. In particular,  the 2-Higgs Doublet 
Model
(2HDM) type III
(henceforth, 2HDM-III for short) provides one with a phenomenology which can be very rich as far as flavor-violating Higgs bosons decays are concerned \cite{Mor, Hernandez, Das}, since, on the one hand, dangerous Flavor Changing Neutral Currents (FCNCs) are controlled by the approach of four-zero texture, which, on the other hand, also enables (amongst others) the above $h\mu^\pm\tau^\mp$ interactions. In fact, motivated by the enhancement of flavor-violating quarks decays $(\phi \to b\bar{s})$ of intermediate mass Higgs bosons (i.e., with mass below the top-quark mass), we focus on the feasibility of  observing the aforementioned LFV processes at the upcoming Large Hadron electron Collider (LHeC) at CERN, which is a future Deep Inelastic Scattering (DIS) experiment at the TeV scale, with center-of-mass energy around 1.3 TeV and  an integrated luminosity of approximately $100$ fb$^{-1}$/year and with a high detector coverage. In this work we study the feasibility of finding, at the next LHeC, the two CP-even neutral Higgs bosons of the 2HDM-III, one being the SM-like Higgs ($h$) and the second one an heavier Higgs state ($H$), both decaying to  $s\bar{b}$ pairs.

\section{The Higgs-Yukawa sector in the 2HDM-III}

In the generic 2HDM, 
the Yukawa Lagrangian is given by
\begin{equation}
  \mathcal{L}_{Y} = -\left(Y_{1}^{u}\bar{Q}_{L}\tilde{\Phi}_{1} u_{R} + Y_{2}^{u}\bar{Q}_{L}\tilde{\Phi}_2 u_{R} + Y_{1}^d \bar{Q}_{L}\Phi_1 d_{R} + Y_2^d\bar{Q}_{L} \Phi_2 d_{R} + Y_{1}^l \bar{L}_L \Phi_1 l_{R} + Y_2^l\bar{L}_L \Phi_2 l_R\right),
\end{equation}
where $\Phi_{1,2}=\left(\phi_{1,2}^+,\phi_{1,2}^0\right)^T$ refer to the 2  Higgs doublets, $\tilde{\Phi}_{1,2} = i \sigma_2\Phi_{1,2}^*$. Besides, the fermion mass matrices after Electro-Weak Symmetry Breaking (EWSB) is given by $M_f = \frac{1}{\sqrt{2}}\left(v_1 Y_1^f + v_2 Y_2^f\right)$, $f=u,d,l$, assuming that both Yukawa matrices $Y_1$ and $Y_2$ have a four-texture form and are Hermitian \cite{Mor}. Taking into account the diagonalization of mass matrices in the following way, $\bar{M}_f = V_{fL}^{\dagger}M_f V_{fR}$, one can obtain  for the rotated matrix $\tilde{Y}_n^q$ the following generic form \cite{Felix}:
\begin{equation}
  \left(\tilde{Y}_{n}^q\right)_{ij} = \frac{\sqrt{m_i^q m_j^q}}{v}\left(\tilde{\chi}_n^q\right)_{ij} = \frac{\sqrt{m_i^q m_j^q}}{v}\left(\chi_n^q\right)_{ij} e^{i\theta_{ij}^q},
\end{equation} 
where the $\chi$'s are unknown dimensionless parameters of the model. One can then have the generic expression for the couplings of the Higgs bosons with the fermions, as follows \cite{Mor, Hernandez}:
\begin{eqnarray}
  \mathcal{L}^{\bar{f}_if_j\phi} & = & -\left\{\frac{\sqrt{2}}{v}\bar{u}_i \left(m_{d_j} X_{ij} P_R + m_{u_i} Y_{ij} P_L \right) d_j H^{+} + \frac{\sqrt{2}m_{l_j}}{v}Z_{ij}\bar{\nu}_L l_R H^{+} + h.c.\right\} \nonumber \\ 
  & & - \frac{1}{v}\left\{\bar{f}_i m_{f_i} h_{ij}^f f_j h^0 + \bar{f}_i m_{f_i} H_{ij}^f f_j H^0 - i\bar{f}_i m_{f_i} A_{ij}^f f_j \gamma_5 A^0 \right\}. 
\end{eqnarray}
where $\phi_{ij}^f$ ($\phi=h,H,A$), $X_{ij},Y_{ij}$ and $Z_{ij}$ are defined in \cite{Mor, Hernandez, Cordero}.
In this work, we consider four different realizations of the 2HDM-III, each enabling an enhancement in the Higgs decay with flavor violation $\phi\to s\bar{b}$. The following benchmark scenarios are in fact studied (Higgs and flavor physics constraints are considered, see details in \cite{Mor, Das,Cordero}):
\begin{itemize}
  \item \textbf{Scenario Ia}: 
$\cos(\beta - \alpha) = 0.1, \; \chi_{kk}^u = 1.5 \; (k=2,3), \; \chi_{22}^d=1.8, \; \chi_{33}^d=1.2, \; \chi_{23}^{u,d}=0.2, \; \chi_{22}^l=0.5,\; \chi_{33}^l=1.2,\; \chi_{23}^l=0.1,\; m_A=100 \, \text{GeV},\; m_{H^{\pm}}=110 ~{\text{GeV}},$ taking $Y=-X=-Z=\cot\beta=2,15,30$.
  \item \textbf{Scenario Ib}: the same as scenario Ia but with $\cos(\beta - \alpha)=0.5$.
  \item \textbf{Scenario IIa}: 
$\cos(\beta - \alpha) = 0.1, \; \chi_{22}^u=0.5, \; \chi_{33}^u=1.4, \; \chi_{22}^{d}=2, \; \chi_{33}^d=1.3,\; \chi_{23}^u=-0.53,\; \chi_{23}^d=0.2,\; \chi_{22}^l=0.4,\; \chi_{33}^l=1.2,\; \chi_{23}^l=0.1,\; m_A=100 ~{\text{GeV}},\; m_{H^{\pm}}=110 ~{\text{GeV}},$ taking $X=Z=1/Y=\tan\beta=2,15,30$.
  \item \textbf{Scenario Y}: 
$\cos(\beta - \alpha) = 0.1, \; \chi_{22}^u=0.5, \; \chi_{33}^u=1.4, \; \chi_{22}^{d}=2, \; \chi_{33}^d=1.3,\; \chi_{23}^u=-0.53,\; \chi_{23}^d=0.2,\; \chi_{22}^l=0.4,\; \chi_{33}^l=1.1,\; \chi_{23}^l=0.1,\; m_A=100 ~{\text{GeV}},\; m_{H^{\pm}}=110 ~{\text{GeV}},$ taking $X=1/Y=-1/Z=\tan\beta=2,15,30$.
\end{itemize}

\section{Numerical Analisys}
For production, we consider the Leading Order (LO) process $ep \to \nu_e \phi q_f$, where $\phi=h,H$, $q$ is a light flavor quark ($u,d,s,c$) whereas $\phi$ decays dominantly into $b \bar{s}$ and the charged conjugation channel (henceforth, $bs$ for short).
Taking into account the different 2HDM-III scenarios mentioned above, we are interested in looking for signal Higgs events in the $bs$ mode. We consider for the electron beam energy $E_e=60$ GeV and for the proton beam $E_p=7000$ GeV, 
yielding a center of mass energy $\sqrt{s}=1.3 TeV$, this in presence of an integrated luminosity of $100$ fb$^{-1}$.
Tab. \ref{tab:sigmabr} shows the model parameters in 2HDM-III for the scenarios Ia, Ib, IIa and Y and the ensuing
signal rates both at production and Branching Ratio (BR) level. For all our benchmarks we have that $\sigma.bs$ is greater than 0.15 fb, so that at least 15 events are produced for an integrated luminosity of 100 fb$^{-1}$.
\begin{table}
 \begin{center}
{\footnotesize
\begin{tabular}{|c|c|c|c|c|c|c|c|c|c|c|c|c|} \hline \hline 2HDM-III & $\tan\beta$ & $X$ &$ Y$ & $Z$ & \multicolumn{2}{c|}{$m_h=125$ GeV} & \multicolumn{2}{c|}{$m_H=130$ GeV} & \multicolumn{2}{c|}{$m_H=150$ GeV} & \multicolumn{2}{c|}{$m_H=170$ GeV} \\ &&&&& bs & $\sigma$.bs & bs & $\sigma$.bs & bs & $\sigma$.bs & bs & $\sigma$.bs \\ \hline \hline   Ia2 & 2 &&&& 0.76 & 0.29 & 0.75 & 0.330 & 0.22 & 0.077 & 0.011 & 0.003 \\
  Ia15 & 15 & $-\cot\beta$ & $\cot\beta$ & $-\cot\beta$ & 12.0 & 11.7 & 0.71 & 0.006 & 0.58 & 0.004 & 0.20 & 0.001 \\
  Ia30 & 30 &&&& 12.8 & 19.1 & 3.16 & 0.088 & 2.50 & 0.027 & 0.80 & 0.005 \\ \hline
  Ib2 & 2 &&&& 0.76 & 0.30 & 0.75 & 0.33 & 0.22 & 0.077 & 0.011 & 0.003 \\
  Ib15 & 15 & $-\cot\beta$ & $\cot\beta$ & $-\cot\beta$ & 8.6 & 7.6 & 23.6 & 5.16 & 8.34 & 1.39 & 0.49 & 0.065 \\
  Ib30 & 30 &&&& 10.9 & 11.5 & 25.2 & 7.5 & 16.9 & 3.18 & 1.85 & 0.240 \\ \hline
  IIa2 & 2 &&&& 0.008 & 0.007 & 15.6 & 0.17 & 4.68 & 0.033 & 0.58 & 0.003 \\\
  IIa15 & 15 & $\tan\beta$ & $\cot\beta$ & $\tan\beta$ & 0.48 & 0.41 & 13.1 & 0.14 & 12.6 & 0.090 & 8.84 & 0.046 \\
  IIa30 & 30 &&&& 2.34 & 1.97 & 13.1 & 0.14 & 13.1 & 0.092 & 11.7 & 0.061 \\ \hline
  Y2 & 2 &&&& 1.33 & 1.12 & 2.62 & 0.026 & 1.90 & 0.013 & 0.50 & 0.0026 \\
  Y15 & 15 & $\tan\beta$ & $\cot\beta$ & $-\cot\beta$ & 0.29 & 0.24 & 20.2 & 0.220 & 4.94 & 0.034 & 0.57 & 0.0030 \\
  Y30 & 30 &&&& 3.98 & 3.36 & 46.8 & 0.518 & 46.0 & 0.336 & 39.2 & 0.2071 \\ \hline \hline
  \end{tabular}
}
\caption{The model parameters for our benchmark points in the 2HDM-III scenario labelled Ia, Ib, IIa, and Y. Here, $bs$ is  the BR$(h,\,H\rightarrow b\bar{s} + \bar{b}s)$, where $\phi=h,H$, in units of $10^{-2}$ while $\sigma.bs$ is the 
production cross-section multiplied by the BR at the LHeC in units of fb.} \label{tab:sigmabr}
 \end{center}
\end{table} 
To estimate the event rate at parton level,  we applied the following basic pre-selections:
\begin{equation} \label{presel}
 p^q_T > 15 \; \text{GeV}, \; \Delta R(q,q) > 0.4,
\end{equation}
where $\Delta R = \Delta \eta^2 + \Delta\phi^2$, $\eta$ and $\phi$ being  the pseudo-rapidity and azimuthal angle, respectively. We passed the CalcHEP v3.4.7 \cite{Belyaev:2012qa} generated parton level event on to PYTHIA v.6.408 \cite{Sjostrand:2006za}, which handles the parton shower, hadronization, heavy hadron decays, etc. We also took the experimental resolutions of the jet angles and energy using the toy calorimeter  PYCELL, in accordance with the LHeC detector parameters given in PYTHIA.
 There are mainly two types of backgrounds for the Higgs signal: charged current backgrounds: $\nu t\bar{b}$, $\nu b \bar{b} j$, $\nu b2j$, $\nu 3j$ and the photo-production channels  $e^- b \bar{b} j$, $e^- t \bar{t}$.
We adopted a simple cut-based method for signal enhancement and background rejection. We have chosen the following selections, and applied these cumulatively, for signals coming from $h(H)$:
\begin{itemize}
 \item \textbf{a(A)}: We first selected that the event must contain at least three jets. 
 \item \textbf{b(B)}: We required at least one $b$-tagged jet with the inclusion of proper mis-tagging. 
 \item \textbf{c(C)}: We required at least two central jets, with $p_T>30$ GeV and $25$($20$)[$15$] GeV for $m_H=130(150)$[$170$] GeV, in the pseudo-rapidity range $|\eta|<2.5$. One of the central jet must be a $b$-tagged jet and we required only one $b$-tagging per event. 
 \item \textbf{d(D)}: The missing transverse energy cut $\slashed{E}_T>20$ GeV is then applied. 
This selection is crucial to suppress the photon production processes.
 \item \textbf{e(E)}: A lepton ($e$ or $\mu$) veto with $p_T>20$ GeV and $\eta<3.0$ is applied. 
 \item \textbf{f(F)}: In the central region, defined above in c(C), we reconstructed the invariant mass of one $b$-tagged jet with all other jets, $M_{bj}$. We have chosen the best combination where the absolute difference $|M_{bj}-M_{h(H)}|$ is minimal within a 15 GeV mass window.
 \item \textbf{g(G)}: We required the remaining leading jet in the event with $p_T>25$ GeV, with $-5.5<\eta<-0.5(-1.0)$
for $h(H)$ and called this the forward tagged jet ($j_f$).
 \item \textbf{h(H)}: The invariant mass of the Higgs boson candidate jets plus the forward tagged jet, which is in fact the overall energy scale of the hard scattering, is such that $m_{h j_f} (m_{H j_f})>190$ GeV.
 \item \textbf{i(I)}: Finally we required only one light flavor jet in the central region. This selection is called the central jet veto and plays an important role in these processes. 
\end{itemize}
For more detail see Ref. \cite{Das}.
\begin{table}
 \begin{center}
  {\tiny {\scriptsize
  \begin{tabular}{|c||c||c|c|c|c|c|c|c|c|c||c|}
   \hline &&&&&&&&&&& \\ Proc&RawEvt& a & b & c & d & e & f & g & h & i & $\Sigma$ \\
   \hline \hline Ia2 & 29.9 & 21.1 & 8.3 & 5.4 & 4.6 & 4.4 & 1.8 & 1.5 & 1.3 & 0.8 & 0.06 (0.19) \\
   Ia15 & 1166.3 & 814.3 & 320.2 & 207.9 & 173.0 & 166.6 & 67.3 & 56.6 & 44.2 & 27.7 & 2.12 (6.7) \\
   Ia30 & 1911.1 & 1294.7 & 539.0 & 342.8 & 282.7 & 274.6 & 102.5 & 78.7 & 46.6 & 29.3 & 2.24 (7.1) \\
   \hline Ib2 & 30.0 & 21.0 & 8.1 & 5.4 & 4.5 & 4.3 & 1.8 & 1.5 & 1.3 & 0.8 & 0.06 (0.19) \\
   Ib15 & 761.5 & 521.0 & 212.5 & 137.5 & 113.3 & 109.6 & 42.1 & 33.5 & 23.2 & 15.0 & 1.15 (3.6) \\
   Ib30 & 1145.3 & 776.2 & 323.1 & 206.8 & 170.6 & 165.3 & 63.3 & 48.6 & 29.5 & 18.8 & 1.44 (4.55) \\
   \hline IIa15 & 40.6 & 28.6 & 11.1 & 7.3 & 6.1 & 5.9 & 2.3 & 2.0 & 1.7 & 1.1 & 0.08 (0.25) \\
   IIa30 & 197.0 & 139.3 & 53.9 & 35.9 & 30.0 & 28.9 & 11.6 & 10.0 & 8.4 & 5.2 & 0.39 (1.23) \\
   \hline Y2 & 112.2 & 79.0 & 30.5 & 20.1 & 16.9 & 16.3 & 6.4 & 5.5 & 4.6 & 2.9 & 0.22 (0.69) \\
   Y15 & 24.2 & 17.0 & 6.6 & 4.4 & 3.7 & 3.5 & 1.4 & 1.2 & 1.0 & 0.6 & 0.05 (0.15) \\
   Y30 & 336.0 & 237.7 & 92.8 & 61.7 & 52.1 & 50.2 & 20.1 & 17.1 & 14.4 & 9.2 & 0.70 (2.2) \\
   \hline \hline $\nu t \bar{b}$ & 50712.1 & 28338.4 & 15293.7 & 9845.0 & 8144.2 & 7532.7 & 2982.1 & 2058.0 & 652.2 & 139.6 & \\
   $\nu b \bar{b} j$ & 14104.6 & 6122.8 & 3656.7 & 1858.5 & 1787.1 & 1650.1 & 257.5 & 152.5 & 82.5 & 15.1 & \\
   $\nu b 2j$ & 18043.1 & 8389.2 & 3013.0 & 1691.5 & 1445.5 & 1373.7 & 389.5 & 206.1 & 77.2 & 11.3 & $B=$170.8 \\
   $\nu 3j$ & 948064.2 & 410393.4 & 15560.9 & 0.0 & 0.0 & 0.0 & 0.0 & 0.0 & 0.0 & 0.0 & $\sqrt{B}=$13.1 \\
   $e b\bar{b} j$ & 256730.1 & 55099.8 & 36353.6 & 12659.8 & 1432.0 & 200.7 & 54.1 & 24.8 & 18.0 & 4.5 & \\
   $e t\bar{t}$ & 783.3 & 685.0 & 384.5 & 265.9 & 179.3 & 26.2 & 11.6 & 10.5 & 3.9 & 0.3 & \\ \hline
  \end{tabular}}}
  \caption{Expected number of events after different combinations of cuts for signal and backgrounds at the LHeC with 100 fb$^{-1}$ integrated luminosity for $m_h =125$ GeV. The number in the final column represent the significance for 1000 fb$^{-1}$.}
\label{tab:h125}
 \end{center}
\end{table}

\begin{table}
 \begin{center}
  {\tiny {\scriptsize
  \begin{tabular}{|c||c||c|c|c|c|c|c|c|c|c|}
\hline &&&&&&&&&& \\ Proc&RawEvt& A & B & CD & E & F & G & H & I &$\Sigma$ \\
 \hline \hline Ia2 & 32.8 & 23.6 & 9.2 & 6.1 & 5.8 & 2.0 & 1.7 & 1.5 & 0.9 & 0.07(0.22)\\
Ib2 &32.8 & 23.7 & 9.2 & 6.1 &5.8  &2.1 & 1.7 & 1.5 & 1.0  & 0.08(0.25)  \\
Ib15 &516.0 &370.0  & 145.0  & 94.7  & 90.9   &30.3 &24.6  & 21.1 &13.5  & 1.11(3.5)  \\
Ib30 & 750.9 &520.6  &210.7  & 143.2 &129.2  & 42.8 & 31.2 & 23.1  & 14.2 & 1.17(3.7)\\
\hline IIa2 &16.7 & 11.8 &4.8  & 3.1 & 3.0 &0.9 & 0.7 & 0.5 & 0.3 &0.02(0.06)\\
\hline Y15 & 22.0 & 15.4 & 6.1 & 3.9 & 3.7 &1.3 & 0.9 &0.7  & 0.5 & 0.04(0.12)\\
Y30 &51.8 &36.3  &14.8  & 9.7 & 9.3 & 3.0 & 2.2 &1.6  &1.1  & 0.09(0.28)\\ 
\hline  $\nu t \bar{b} j$&50712.1 & 28338.4 & 15293.7 & 9092.4 & 8393.6 & 2550.9 & 1565.5.8 & 617.9 & 113.7\\
   $\nu b \bar{b} j$ &14104.6 & 6122.8 & 3656.7  & 2062.1 &  & 199.3 & 112.4 &70.8 &12.4\\
   $\nu b 2j$ &18043.1 & 8389.2 & 3013.0 &1734.0 & 1650.1 & 402.8 & 143.7 & 64.5 & 8.1  & $B=$147.8 \\
   $\nu 3j$ & 948064.2 & 410393.4 &15560.9 & 0.0 & 0.0 & 0.0 & 0.0 & 0.0 &0.0& $\sqrt{B}=$12.2 \\
   $e b\bar{b} j$ & 256730.1 & 55099.8 & 36353.6 & 1826.6 & 284.1 & 56.4 & 31.6 & 22.6 & 11.3 \\
   $e t\bar{t}$ &783.3 & 685.0 & 384.5 & 190.8 & 27.8 & 10.9 & 9.3 & 3.9 &0.3 \\ \hline
\end{tabular}}}
\caption{The same as Tab. 2, but for $m_H=130$ GeV.}
\label{tab:H130}
\end{center}
\end{table}

\begin{table}
 \begin{center}
  {\tiny {\scriptsize
  \begin{tabular}{|c||c||c|c|c|c|c|c|c|c|c|}
\hline &&&&&&&&&& \\ Proc&RawEvt& A & B & CD & E & F & G & H & I &$\Sigma$ \\
 \hline \hline Ib15 &139.6 & 108.2 & 41.7 & 31.6 & 29.9 & 7.0 & 5.9 & 5.3 & 3.7 & 0.48(1.5)\\
IIb30 & 317.6 & 234.5 & 91.5 & 68.6 & 65.2 & 14.7 & 11.7 & 10.5 & 7.4 & 0.95(3.0)\\
\hline Y30 & 33.6 & 25.3 & 9.9 & 7.5 & 7.1 & 1.7 & 1.4 & 1.2 & 0.9 & 0.12(0.38)\\
\hline  $\nu t \bar{b} j$&50712.1 & 28338.4 & 15293.7 & 9808.7 & 9039.0 & 751.7 & 476.8 & 194.5 & 32.3\\
   $\nu b \bar{b} j$ &14104.6 & 6122.8 & 3656.7  & 2300.1 & 2120.8 & 199.3 & 112.4 &70.8 &12.4\\
   $\nu b 2j$ &18043.1 & 8389.2 & 3013.0  & 2030.3 & 1933.1 & 234.2 & 83.7 & 41.0 & 6.3 & $B=$60.1 \\
   $\nu 3j$ & 948064.2 & 410393.4 &15560.9 & 0.0 & 0.0 & 0.0 & 0.0 & 0.0 &0.0& $\sqrt{B}=$7.7 \\
   $e b\bar{b} j$ & 256730.1 & 55099.8 & 36353.6 & 2270.8 & 385.6 & 36.1 & 24.8 & 20.3 & 9.0\\
   $e t\bar{t}$ &783.3 & 685.0 & 384.5 & 199.0 & 29.1 & 3.5 & 3.0 & 1.2 & 0.1\\ \hline
\end{tabular}}}
\caption{The same as Tab. 2, but for $m_H=150$ GeV.}
\label{tab:H150}
\end{center}
\end{table}
Tab 2. shows the expected number of events after different combinations of cuts for signal and backgrounds at the LHeC with 100 fb$^{-1}$ integrated luminosity for $m_h$=125 GeV. 
 RawEvt stands for the number of events with only generator level cuts imposed, for signal as well as for background:
 these are calculated from the total cross-section times BR. In the final column we mention the significances $(\Sigma)$ defined as $S=S/\sqrt{B}$, where $S(B)$ are  signal(background) events for 100 fb$^{-1}$ of data and the number between parenthesis represents significances for 1000 fb$^{-1}$. Tabs. 3 and 4 show the same, but now for the heaviest neutral Higgs $H$, taking $m_H=130$ and 150 GeV, respectively.  One can see that, for 100 fb$^{-1}$ of data,  $h$ would be observed in the LHeC experiment  in the following scenarios: Ia, Ib,  with approximately $(1-2)\Sigma$. The heavier neutral Higgs boson, $H$, with a mass 150 GeV would have a $1\Sigma$ significance  in scenario Ib. At the end of the
LHeC operation, an accumulated total integrated luminosity of 1000 fb $^{-1}$ is expected,  which would allow for the  detection of both Higgs bosons.

 \section{Conclusions}

At the LHeC, having  considered 100 fb$^{-1}$ of data, we found that the SM-like Higgs boson, $h$, could be observed  with approximately $(1-2)\Sigma$ significance. The heavier neutral Higgs boson, $H$, with a mass 150 GeV would have $1\Sigma$ significance. At the end of the LHeC lifetime, with  1000 fb$^{-1}$, the detection of both Higgs bosons ($h, H$) would be certain. All this would occur in the quark-flavor violating 
decay channels $h,H\to b\bar s$ (plus charge conjugate), which would then serve as a probe of new physics, which
we have herein attributed to a 2HDM-III.

\end{document}